\documentclass[%
 reprint,
 amsmath,amssymb,
 aps,
]{revtex4-1}
\usepackage{amsthm}
\usepackage{graphicx}
\usepackage{float}
\usepackage{dcolumn}
\usepackage{bm}
\usepackage{extarrows}
\usepackage{hyperref}
\usepackage{float}
\usepackage{color}

\usepackage{marvosym}
\usepackage{epstopdf}
\usepackage{subfigure}
\usepackage{array}
\usepackage{amssymb}
\usepackage{varioref}
\usepackage[utf8]{inputenc}
\usepackage{indentfirst}
\begin{document}

\preprint{APS/123-QED}

\title{Quantum Alternating Operator Ansatz for Solving the Minimum Exact Cover Problem}

\author{Sha-Sha Wang $^{1}$}

\author{Hai-Ling Liu $^{1}$}

\author{Yan-Qi Song $^{1}$}

\author{Su-Juan Qin $^{1}$}

\author{Fei Gao $^{1}$}
\email{gaof@bupt.edu.cn}

\author{Qiao-Yan Wen $^{1}$}
\email{wqy@bupt.edu.cn}

\affiliation{$^{1}$ State Key Laboratory of Networking and Switching Technology, Beijing University of Posts and Telecommunications, Beijing 100876, China}
\date{\today}

\begin{abstract}
The Quantum Alternating Operator Ansatz (QAOA+) is an extension of the Quantum Approximate Optimization Algorithm (QAOA), where the search space is smaller in solving constrained combinatorial optimization problems. However, QAOA+ requires a trivial feasible solution as the initial state, so it cannot be used for problems that are difficult to find a trivial feasible solution. For simplicity, we call them as Non-Trivial-Feasible-Solution Problems (NTFSP). In this paper, we take the Minimum Exact Cover (MEC) problem as an example, studying how to apply QAOA+ to NTFSP. As we know, exact covering (EC) is the feasible space of MEC problem, which has no trivial solutions. To overcome the above problem, the EC problem is divided into two steps to solve. First, disjoint sets are obtained, which is equivalent to solving independent sets. Second, on this basis, the sets covering all elements (i.e., EC) are solved. In other words, we transform MEC into a multi-objective constrained optimization problem, where feasible space consists of independent sets that are easy to find.  Finally, we also verify the feasibility of the algorithm from numerical experiments. Our method provides a feasible way for applying QAOA+ to NTFSP, and is expected to expand its application significantly.
\end{abstract}

\pacs{Valid PACS appear here}
\maketitle


\section{Introduction}
Quantum computers have computational advantages over classical computers by exploiting quantum effects, providing polynomial or even exponential speedups for specific problems, such as integer factorization \cite{1}, unstructured data search \cite{2}, linear regression \cite{4,5}, dimension reduction \cite{7,8,9,10}, quantum error correction \cite {3}, matrix computation \cite{11,12,13,14}, anomaly detection \cite{15} and cryptanalysis \cite{16}. However, the current quantum hardware devices only support a limited number of physical qubits and limited gate fidelity, which makes the above quantum algorithms unable to be implemented on near-term devices.

Recently, Quantum Approximation Optimization Algorithm (QAOA) \cite{17} is a kind of hybrid quantum-classical algorithm, which can be implemented on Noisy Intermediate-Scale Quantum (NISQ) device \cite{18}. QAOA has successfully solved many combinatorial optimization problems, such as max cut \cite{19,20}, minimum vertex cover \cite{21}, and correlation clustering \cite{22}.

In 2019,  Hadfield et al. presented the Quantum Alternating Operator Ansatz (QAOA+) \cite{31} to solve the combinatorial optimization problems. The algorithm framework is similar to QAOA, and the circuit consists of $p$ alternating layers of  the phase-separation operator $U(H_P, \gamma)$ and the mixing operator $U(H_M, \beta)$ applied to an initial state $|x\rangle$. Different from QAOA, $U(H_M, \beta)$ of QAOA+ will preserve the feasible space (the space composed of feasible solutions), and the initial state $|x\rangle$ is required to be a trivial feasible state (the quantum state corresponding to a trivial feasible solution). The algorithm can limit the state of the system to the feasible space, resulting in zero probability of obtaining invalid solutions, which implies a prominent advantage compared to QAOA. QAOA+ has been applied to many combinatorial optimization problems, such as  graph-coloring \cite{33}, maximum independent set \cite{34}, max-$k$ vertex cover \cite{35}, graph matching \cite{36}, and lattice protein folding \cite{37}. However, the application of QAOA+ has a prerequisite that it needs to be easy to find a trivial feasible solution of the problem (as the initial state).  QAOA+ is not suitable for problems which is difficult to find a trivial feasible solution. For simplicity, we call the above problems as Non-Trivial-Feasible-Solution Problems (NTFSP). In this paper, we will take the MEC problem as an example, trying to apply QAOA+ to NTFSP.

\raggedbottom The Minimum Exact Cover (MEC) is a constrained optimization problem, with wide applications in the tail-assignment and vehicle routing. Some scholars solved this problem using QAOA \cite{23,24,25} and achieved good results. However, because Exact Cover (EC) has no trivial solutions, there is no relevant research to solve MEC using QAOA+.

In this paper, taking MEC problem as an example, we study how to apply QAOA+ to NTFSP. Specifically, according to the describe of EC problem \cite{25}, it can be found intuitively that elements of an EC are disjoint sets, noted as $S^* = \{S_l, \cdots , S_k\}$, $2 \le l, k \le n $ ($n$ is the number of qubits), and $\left| S_l \right|  + \cdots + \left| S_k \right| = m$ ($m$ is the number of all elements) can be obtained. Therefore, to create a trivial feasible initial state $|x\rangle$, we divide the EC problem into two steps to solve. First, inspired by \cite{30}, we construct a graph in which each set $S_i(i=1, \cdots, n)$ is regarded as a vertex. And, there is an edge between two vertices $S_i$, $S_j$ if and only if $S_i \cap S_j \neq \emptyset$. It is found that disjoint sets are solved, which is equivalent to obtaining the independent sets on the graph. Second, on this basis, the sets covering all elements (i.e., EC) are solved, which is equivalent to maximizing $\left| S_l \right|  + \cdots + \left| S_k \right|$. In other words, we transform MEC into a multi-objective constrained optimization problem, where feasible solutions are independent sets that are easy to find.  To solve the above multi-objective constrained optimization problem, we adopt the linear weighted sum method to construct target Hamiltonian. Finally, we perform numerical experiments with 6, 8, and 10 qubits by using MindQuantum \cite{51}. The numerical results show that the solution can be obtained with high probability, even though level $p$ of the algorithm is low. Besides, to optimize quantum circuit, we remove single qubit rotating gates $R_Z$. It is found that $p$-level optimized circuit only needs $p$ parameters, which can achieve an experimental effect similar to original circuit with $2p$ parameters.

We apply QAOA+ to solve NTFSP, which provides a meaningful reference for how to use QAOA+ to solve such problems, and might greatly expand the application of the algorithm.

This paper is organized as follows. In Sec. II, QAOA+ is reviewed. In Sec. III, we apply QAOA+ to solve MEC problem. In Sec. IV, numerical results and analysis are given. Finally, the summary and prospects are given in Sec. V.

\section{Review of QAOA+}
The optimization problem $(F,f)$ is considered, where $F$ is the feasible set and $f:$  $F\rightarrow \mathbb{R}$ is the objective function to be optimized. Let $\mathcal{F}$ be the Hilbert space of dimension $|F|$, whose standard basis is $\left\{|x\rangle: x \in F \right\}$. As a kind of hybrid quantum-classical algorithm, QAOA+ \cite{31} is often used to solve combinatorial optimization problems. The specific process of QAOA+ is as follows.

First, we should be able to create an initial state $|x\rangle, x \in F$ which is a trivial feasible solution, or the uniform superposition state of trivial feasible solutions. Second, the phase-separation operator $U(H_P, \gamma) = e^{- i \gamma H_P}$ which depends on the objective function $f$,  and the mixing operator $U(H_M, \beta) = e^{- i \beta H_M}$ which depends on $F$ and its structure are applied alternately to $|x\rangle$, where $\gamma$ and $\beta$ are real parameters, and $H_M$ and $H_P$ are the mixing Hamiltonian and the target Hamiltonian, respectively. It worth noting that $U(H_M, \beta)$ needs to meet two conditions: 1) preserve the feasible subspace; 2) provide transitions between all pairs of feasible spaces, see Ref.\cite{31} for details.

The alternating sequence continues for a total of $p$ times with different variational parameters $\overrightarrow{\gamma} = (\gamma_1, \gamma_2, \cdots, \gamma_p)$ and $\overrightarrow{\beta} = (\beta_1, \beta_2, \cdots, \beta_p)$, where $\gamma_i \in [0, 2\pi]$, $\beta_i \in [0, \pi]$, such that the final variational state becomes
\begin{align}
|\psi_p(\overrightarrow{\gamma}, \overrightarrow{\beta})\rangle = U(H_M, \beta_p)U(H_P, \gamma_p) \cdots U(H_P, \gamma_1)|x\rangle.
\end{align}

The variational parameters are optimized on classical computers. The structure of the QAOA+ is shown in Fig. 1. The objective of the classical optimizer is to find the optimal parameters $(\overrightarrow{\gamma^*}, \overrightarrow{\beta^*})$, which are obtained by maximizing the expected value of the  target Hamiltonian
\begin{align}
  (\overrightarrow{\gamma^*}, \overrightarrow{\beta^*})\ = arg \ \mathop{max}\limits_{\overrightarrow{\gamma}, \overrightarrow{\beta}}\ F_p(\overrightarrow{\gamma}, \overrightarrow{\beta}),
\end{align}
where $F_p(\overrightarrow{\gamma}, \overrightarrow{\beta}) = \langle\psi_p(\overrightarrow{\gamma}, \overrightarrow{\beta})| H_P |\psi_p(\overrightarrow{\gamma}, \overrightarrow{\beta})\rangle.$

\begin{figure*}
\centering
 \includegraphics[width=15cm]{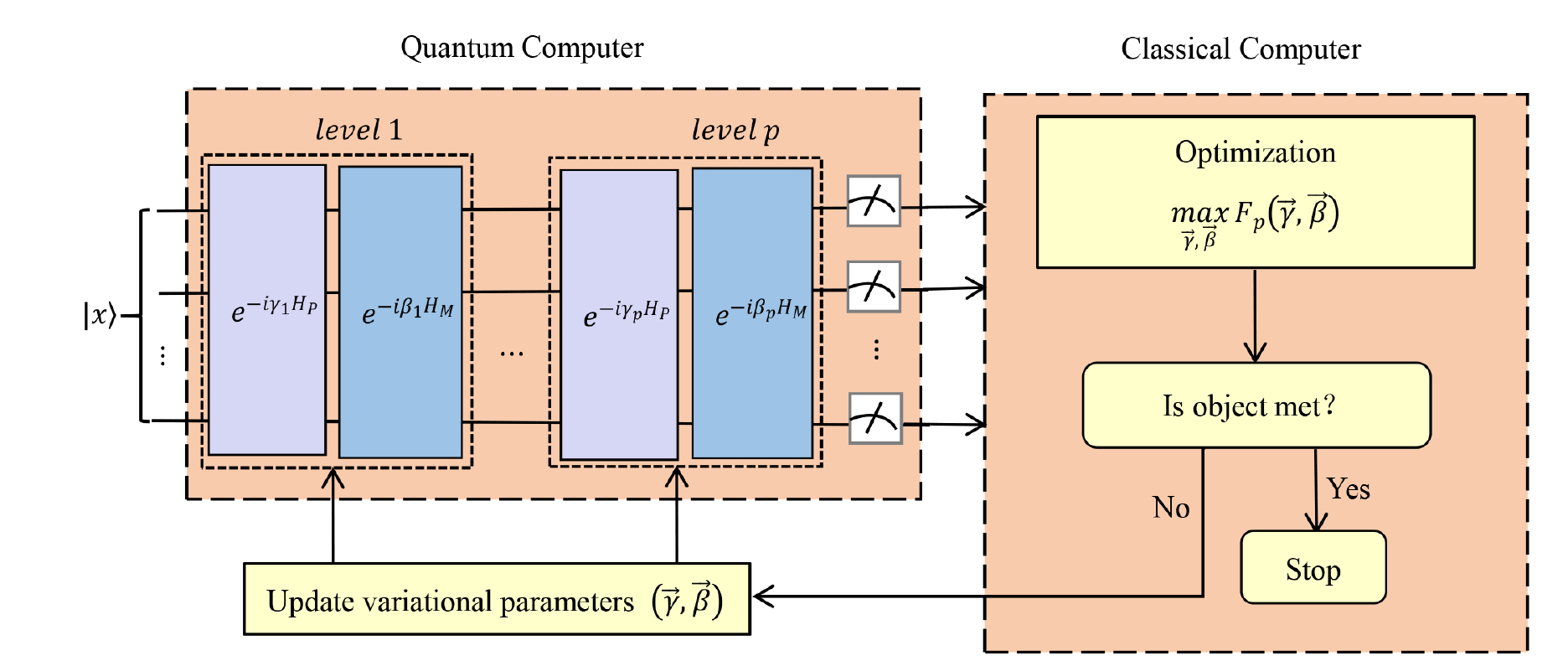}
 \caption{Schematic of the QAOA+ \cite{23}. The quantum processor consists of three parts: initial state, phase-separation operators $U(H_P, \gamma)$, and mixing-operators $U(H_M, \beta)$. The variational parameters are optimized on classical computers. The quantum computer is used to evaluate the expectation value of the objective function.}
 \label{FIG:1}
\end{figure*}

We define the success probability as the probability of finding the optimal solution
\begin{align}
P_{success} = |\langle x_{sol}|\psi_p(\overrightarrow{\gamma}, \overrightarrow{\beta})\rangle|^2,
\end{align}
where $x_{sol}$ is the solution to the problem.

\section{Apply QAOA+ to solve MEC Problem}
\label{Sec:Variational}
In this section, we first introduce MEC. Then, MEC  is transformed into a constrained optimization problem with two objective functions. Finally, we solve MEC  using QAOA+.

\subsection{MEC }
\label{Subsec:Decomposition}
MEC \cite{25} is shown as follows: the sets $X = \{x_1, \cdots , x_m\}$ and $S = \{S_1, \cdots , S_n\}$ are considered, where $S_i \subset X(i = 1, \cdots, n)$, such that
\begin{align}
  X = \bigcup_{i=1}^{n}S_i.
\end{align}

A subset $S^*$ of $S$, it is called an EC of $X$ when elements of $S^*$ are disjoint sets and union of the elements of $S^*$ is $X$.  $S^*$ with the least number of elements is called MEC (MEC is not unique).

MEC problem can be expressed as follows \cite{25}
\begin{align}
& min    \quad  \sum_{i=1}^ns_i,  \label{eq5} \\
& s. t.  \quad  \sum_{i: x_j \in S_i}s_{i} = 1, \quad \forall x_j \in X, \label{eq6}\\
& \quad \quad s_i, s_j \in \{0, 1 \},
\end{align}
where $s_i$ is the label of set $S_i$. When $S_i$ is selected, $s_i=1$, otherwise $s_i=0$. The objective function  Eq. (\ref{eq5}) is to minimize the number of sets, subject to constraints  Eq. (\ref{eq6}) ensuring that the elements of X are covered only once, i.e., EC.

Some algorithms \cite{23,24,25} have been proposed to solve this problem using QAOA. The key point of QAOA is to construct the target Hamiltonian which includes solutions of the MEC problem. For constrained optimization problems, the common method is to incorporate hard constraints into the target function as a penalty item, and then convert the target function into a target Hamiltonian \cite{25,26,27,28,29,30}.  Different from QAOA, $U(H_M, \beta)$ of QAOA+ will preserve the feasible space, and the initial state $|x\rangle$ is required to be a trivial feasible state. The algorithm can limit the state of the system to the feasible space, resulting in zero probability of obtaining invalid solutions, which implies a prominent advantage compared to QAOA. However, EC is the feasible space for the MEC problem, which has no trivial solutions.

According to the describe of EC problem \cite{25}, it can be found intuitively that elements of an EC are disjoint sets, noted as $S^* = \{S_l, \cdots , S_k\}$, $2 \le l, k \le n $ ($n$ is the number of qubits), and $\left| S_l \right|  + \cdots + \left| S_k \right| = m$ ($m$ is the number of all elements) can be obtained. $S^*$ with the least number of elements is called MEC. To create a trivial feasible initial state $|x\rangle$, let $ \omega_i = \left| S_i \right|$, we transform MEC into the following multi-objective constrained optimization problem
\begin{align}
& min    \quad  \sum_{i=1}^ns_i, \label{eq8}\\
& max     \quad \sum_{i=1}^n\omega_is_i, \label{eq9}\\
& s. t.  \quad  s_i + s_j \le 1, S_i \cap S_j \neq \emptyset, \label{eq10}\\
& \quad \quad s_i, s_j \in \{0, 1 \},
\end{align}
where the trivial feasible solutions are easy to find. The objective function Eq. (\ref{eq8}) is to minimize the number of sets and the objective function  Eq. (\ref{eq9}) is to maximize the sum of the weights of sets, subject to constraints  Eq. (\ref{eq10}) ensuring that two sets $S_i, S_j$ cannot be selected simultaneously, where $S_i \cap S_j \neq \emptyset$. Noting that  Eq. (\ref{eq8}) needs to be optimized on the premise of meeting  Eq. (\ref{eq9}).

The linear weighted sum method \cite{38} is the simplest method to solve the multi-objective optimization problem. According to importance of each objective function to determine the corresponding weight, this method transforms the multi-objective into single-objective optimization problem. We adopt the linear weighted sum method to solve the above optimization problem, and transform it into a single-objective optimization problem
\begin{align}
& max     \quad \lambda_1f_1 - \lambda_2f_2, \label{eq12}\\
& s. t.  \quad  s_i + s_j \le 1, S_i \cap S_j \neq \emptyset,\label{eq13}\\
& \quad \quad s_i, s_j \in \{0, 1 \},
\end{align}
where $\lambda_1 > \lambda_2 > 0$, $f_1 = \sum_{i=1}^n\omega_is_i$, $f_2 = \sum_{i=1}^ns_i$.

\subsection{QAOA+ FOR MEC}
\label{Subsec:Evaluate}
  The QAOA+ mapping comprises phase-separation operators $U(H_P, \gamma)$,  mixing operators $U(H_M, \beta)$, and initial state $|x\rangle$ for MEC problem. $U(H_P, \gamma) = e^{-i \gamma H_P}$ depends on  Eq. (\ref{eq12}). $U(H_M, \beta) = e^{-i \beta H_M}$ depends on Eq. (\ref{eq13}) and its structure. According to Eq. (\ref{eq13}), the initial state $|0\rangle^{\bigotimes n}$ can be chosen, which is a trivial feasible solution. Next, we need to construct the mixing Hamiltonian $H_M$ and the target Hamiltonian $H_P$.

To construct $H_M$, inspired by \cite{30}, we construct a graph in which each set $S_i(i=1, \cdots, n)$ is regarded as a vertex. There is an edge between two vertices $S_i$, $S_j$ if and only if $S_i \cap S_j \neq \emptyset$. Therefore, the constraint Eq. (\ref{eq13}) is equivalent to solving independent sets on the graph.

Given an independent set $S^{\prime}$ ($s_i = 1$ if and only if $S_i \in S^{\prime}$, else $s_i = 0$),  to maintain the property of the independent set, the following rules should be met when adding and deleting vertices \cite{31}: 1) the neighbors of $S_i$ are marked as $S_{i1}, S_{i2}, \cdots , S_{il}$, adding a vertex $S_i \notin S^{\prime}$ so that the new point set is still an independent set only if none of the neighbors $S_{i1}, S_{i2}, \cdots , S_{il}$ of $S_i$ are already in $S^{\prime}$, i.e., $s_{i1} = s_{i2} = \cdots = s_{il} = 0$; 2) removing any vertex $S_j \in S^{\prime}$ without affecting the feasibility of the state.  Hence, a bit-flip operation at a vertex, controlled by its neighbors, suffices both to remove and add vertices while maintaining the independence property.

The mixing Hamiltonian $H_M$ is expressed as follows
\begin{align}
H_M = \sum_{i=1}^n\sum_{m_i=1}|s_{i1}s_{i2} \cdots s_{il}\rangle\langle s_{i1}s_{i2} \cdots s_{il}|\otimes X_i,
\end{align}
where $m_i = \prod_{k=1}^l (1-s_{ik})$.

The objective function $f = \lambda_1\sum_{i=1}^n\omega_is_i - \lambda_2 \sum_{i=1}^ns_i$, and target Hamiltonian $H_P$ is obtained by replacing $s_i$ with $\frac{1-\sigma_i^Z}{2}$
\begin{align}
H_P = \lambda_1\sum_{i=1}^n\omega_i\frac{1-\sigma_i^Z}{2} - \lambda_2 \sum_{i=1}^n\frac{1-\sigma_i^Z}{2},
\end{align}
where $\lambda_1>\lambda_2>0$, $\sigma_i^Z$ represents Pauli-$Z$  operation on $i$th qubit.

\section{Numerical Simulation}
We study instances for three different problem sizes of the MEC problem given in Table I, corresponding to 6, 8, and 10 qubits, respectively.
\begin{table}[!ht]
\caption{Information about the MEC problem instances}
\begin{tabular}{|p{20pt}<{\centering}|p{20pt}<{\centering}|c|p{90pt}<{\centering}|}
\hline
$|X|$ &  $|S|$ &  Number of instances & Number of solutions for each instance  \\
\hline
12  & 6 & 10 & 1  \\
\hline
16  & 8 & 10 & 1  \\
\hline
20  & 10 & 10 & 1  \\
\hline
\end{tabular}
\end{table}

\subsection{Discussion about weights}
\label{Subsec:Evaluate}
In multi-objective problems, the weight of each index is one of the important factors that affect the accuracy of the results. Next, we discuss the weight $\lambda_i$ $(i = 1, 2)$ of MEC in two cases and fix $\lambda_1$ and $\lambda_2$.

The MEC is expressed as the following optimization problem
\begin{align}
& max     \quad f, \\
& s. t.  \quad  s_i + s_j \le 1, S_i \cap S_j \neq \emptyset, \\
& \quad \quad s_i, s_j \in \{0, 1 \},
\end{align}
where $\lambda_1 > \lambda_2 > 0$, $f = \lambda_1\sum_{i=1}^n\omega_is_i - \lambda_2 \sum_{i=1}^ns_i$.

Suppose the set $A$ is a MEC (solution of MEC problem), and the corresponding objective function is $f_{A} = \lambda_1m-\lambda_2m^{\prime}$, where $m= | X |$, $m^{\prime}= |A|$. Suppose the set $B$ is not MEC, and its corresponding objective function is $f_{B} = \lambda_1t-\lambda_2t^{\prime}$, where $t\leq m, t^{\prime} = |B|$. Based on the above assumptions, we can obtain $f_{A} >f_{B}$. Next, we will consider that B is an EC or not, and discuss weight $\lambda_i$ in two situations.

If set $B$ is an EC, we can obtain $t = m, t^{\prime}>m^{\prime} $. The inequality $f_{A} >f_{B}$ is always true with $\lambda_i>0$ $(i=1, 2)$. If set $B$ is not an EC, we can get $ t < m $, and then deduce $\frac{\lambda_1}{\lambda_2} > \frac{m^{\prime}-t^{\prime}}{m -t}$.

To determine the appropriate values of $\lambda_1$ and $\lambda_2$, the range of function $f = \lambda_1f_1 - \lambda_2f_2$ is limited to $(0, 1]$
\begin{align}
0<n\lambda_2f_1 - \lambda_2f_2\leq 1.
\end{align}

Because of $nf_1 - f_2>0$, we can deduce $0<\lambda_2\leq\frac{1}{nf_1 - f_2}$, where $f_1\leq m, f_2\geq 2$ (without regard to $S_i$ = X). Further, $0<\lambda_2\leq\frac{1}{nm - 2}$ can be obtained. Without losing generality, we can make $\lambda_2=\frac{1}{nm - 2}$ in this paper. And then, we make $\frac{\lambda_1}{\lambda_2} = n$ according to $\frac{m^{\prime} - t^{\prime}}{m -t} < n$ ($n$ is number of qubits).
\subsection{Low levels: patterns in optimal variational parameters}
\label{Subsec:Evaluate}
The patterns in the optimal variational parameters for MaxCut have been observed in Ref.\cite{39}, where it was found that there is a linear relationship between the parameters and the level $p$. Based on observed linear patterns, two heuristic optimization strategies are proposed, which significantly speed up the classical optimization of QAOA.  The optimal parameter pattern is a useful guide in the selection and design of heuristic strategies. Before studying the performance of the QAOA+, we need to observe the patterns of optimal variational parameters at low level, namely up to $p = 5$.

To find the optimal variational parameters for $1 \leq p \leq 5$,  the gradient-based Broyden-Fletcher-Goldfarb-Shanno (BFGS) \cite{40,41,42,43} is adopted in this paper. It is a commonly used local optimization algorithm, which is repeated with sufficiently many random initial parameters to find the global optimum.

After performing numerical simulations for 10 instances of MEC problem with 6-qubit using QAOA+, we present the optimal variational parameters $(\overrightarrow{\gamma^*}, \overrightarrow{\beta^*})$ from $p=3$ up to $p=5$ as shown in Fig. 2. The optimal parameters do not follow the linear pattern as in Ref.\cite{39} so the interpolation optimization cannot be performed. Recently,  parameters fixing strategy \cite{44}, a straightforward, yet practically effective, is proposed to improve the performance of QAOA at large circuit depths. The optimal parameters $(\gamma_1^*, \cdots,\gamma_{p-1}^*,\beta_1^*, \cdots, \beta_{p-1}^*)$ at level $p-1$ to be further optimized as they are passed into QAOA of $p$-level as the initial parameters. Hence, the initial parameters at level $p$ will be $(\gamma_1^*, \cdots,\gamma_{p-1}^*,\gamma_{p}, \beta_1^*, \cdots, \beta_{p-1}^*,\beta_{p})$, where $\gamma_{p}$ and $\beta_{p}$ are random parameters. Inspired by this, we will investigate whether this strategy can improve the performance of QAOA+.

\begin{figure*}[htp]
\centering
 \includegraphics[width=12cm]{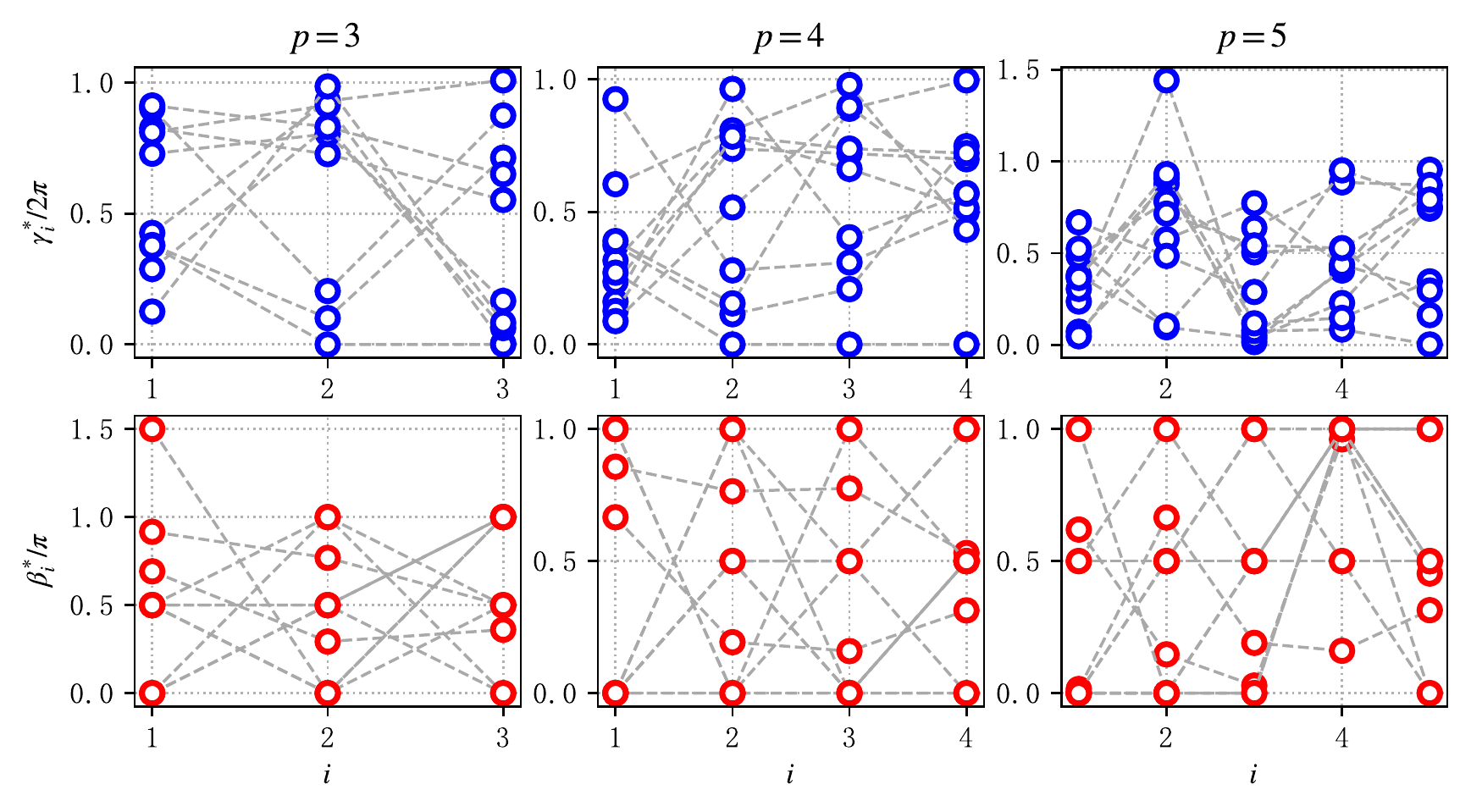}
 \caption{Optimal QAOA+ variational parameters  for 10 instances with 6-qubit, for $3 \leq p \leq 5$. In the above figure, each dashed line connects parameters for one instance. For each instance and each $p$, we use the classical BFGS optimization method from 1000 random initial variational parameters, and keep the best parameters.}
 \label{Fig.2}
\end{figure*}
\renewcommand{\thefigure}{3}
\begin{figure*}[htbp]
	\centering
	\begin{minipage}{0.43\linewidth}
		\centering
		\includegraphics[width=7.4cm]{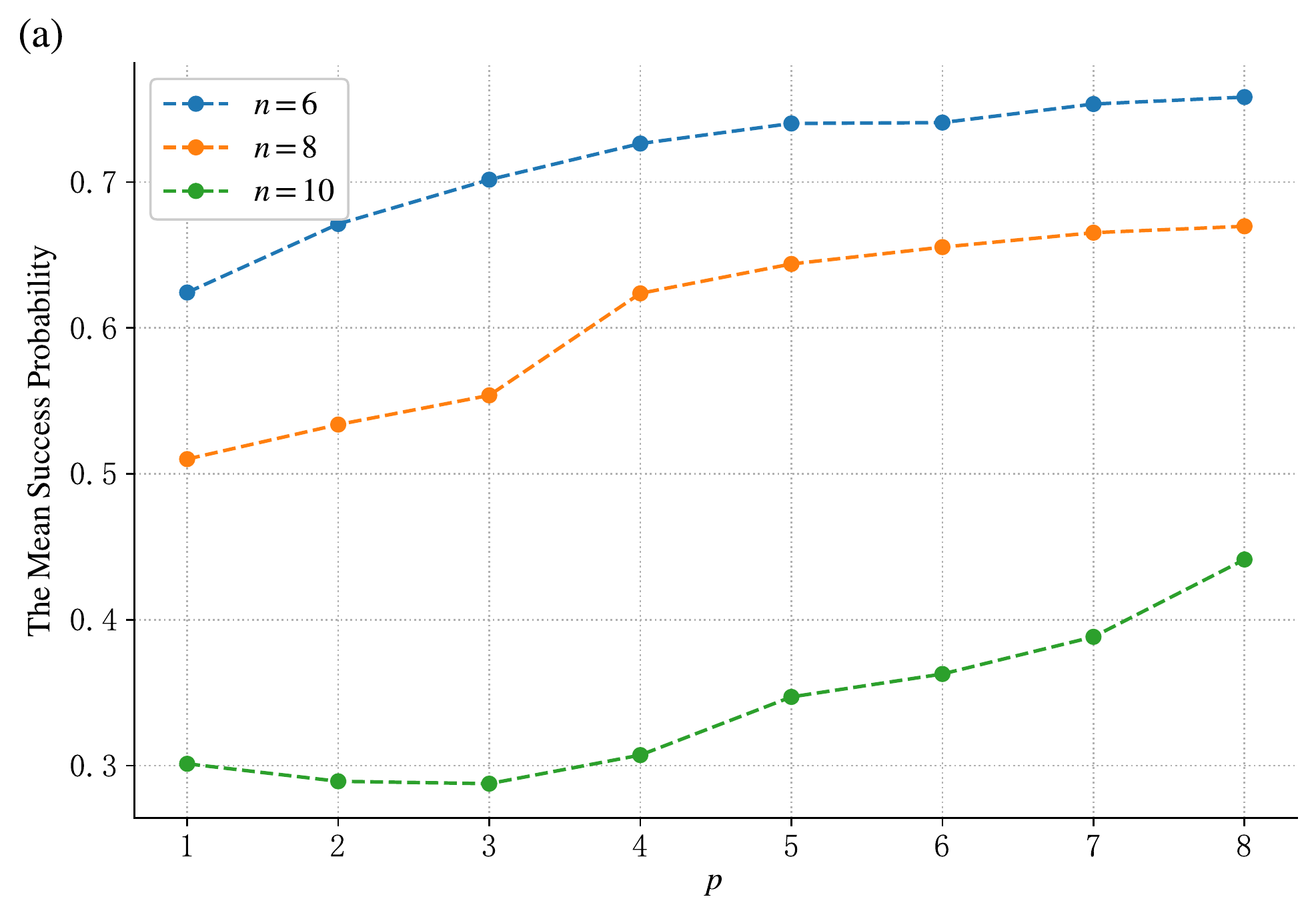}
	\end{minipage}
	\begin{minipage}{0.43\linewidth}
		\centering
		\includegraphics[width=7.4cm]{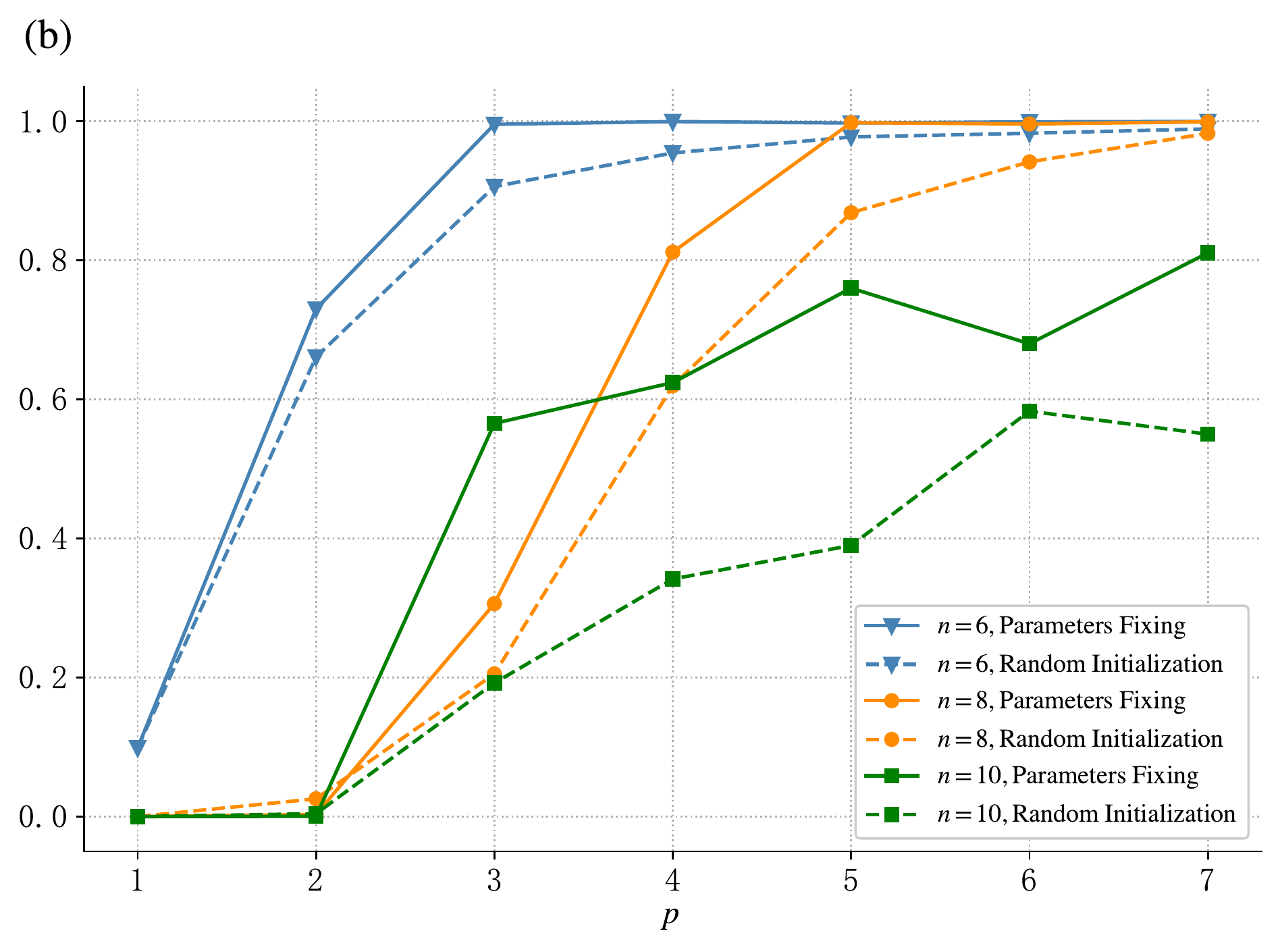}
	\end{minipage}
     \caption{(a) The mean success probability, as a function of level $p$ averaged, is plotted using random initialization method, over all instances for the three different problem sizes. (b) The comparison is drawn between the parameters fixing strategy and the random initialization approach for optimizing the QAOA+, on one selected instance from each problem size.}
\label{Fig.3}
\end{figure*}

\subsection{Analysis of success probability}
\label{Subsec:Evaluate}
Based on the discussion of optimal variational parameter patterns in the previous section, we intend to use random initialization and parameters fixing strategy to study the performance of QAOA+. From the starting point of generation, we run the BFGS optimization method for this algorithm.

In Fig. 3(a), the mean success probability, as a function of level $p$, is plotted with random initialization parameters method, over all instances for the three different problem sizes. It is observed that the mean success probability increased slowly with the increase of level $p$ overall in all 30 instances.

To investigate whether parameters fixing strategy can improve the performance of QAOA+, we select an instance from each problem size, and simulate them up to $p=7$ using random initialization and parameters fixing strategy respectively. In Fig. 3(b), we compare  the parameters fixing strategy to the random initialization approach for optimizing the QAOA+. For random initialization method, it is observed that for 6-qubit and 8-qubit, the solution can be obtained with a probability close to 100\% with $p=7$. For parameters fixing strategy, for 6-qubit and 8-qubit, we obtain a  mean success probability close to 100\% with $p = 3$ and $p = 5$, respectively. In general, the parameters fixing strategy outperforms the random initialization run for these three examples.

The numerical results show that the solution can be obtained with high probability by using parameters fixing strategy, even though level $p$ of the algorithm is low. Hence, we can conclude that the parameters fixing strategy can make the algorithm have better performance, at least true for the MEC problem instances used in our work.

\renewcommand{\thefigure}{5}
\begin{figure*}[htp]
\centering
    \includegraphics[width=14cm]{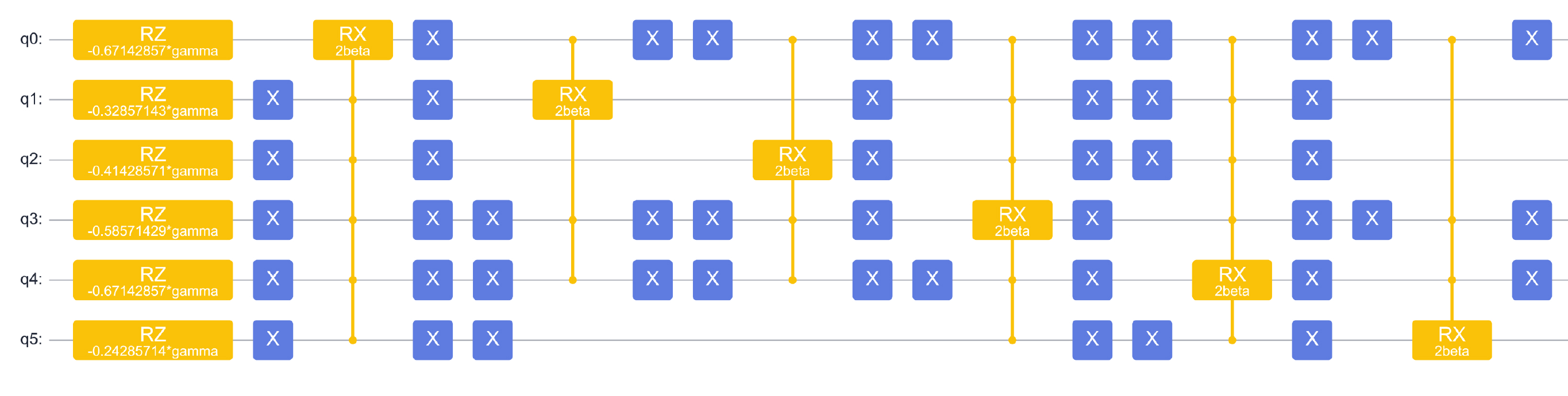}
    \caption{The overall 1-level quantum circuit diagram of QAOA+ mapping.}
\label{Fig.5}
\end{figure*}
\renewcommand{\thefigure}{6}
\begin{figure*}[htp]
\centering
    \includegraphics[width=11cm]{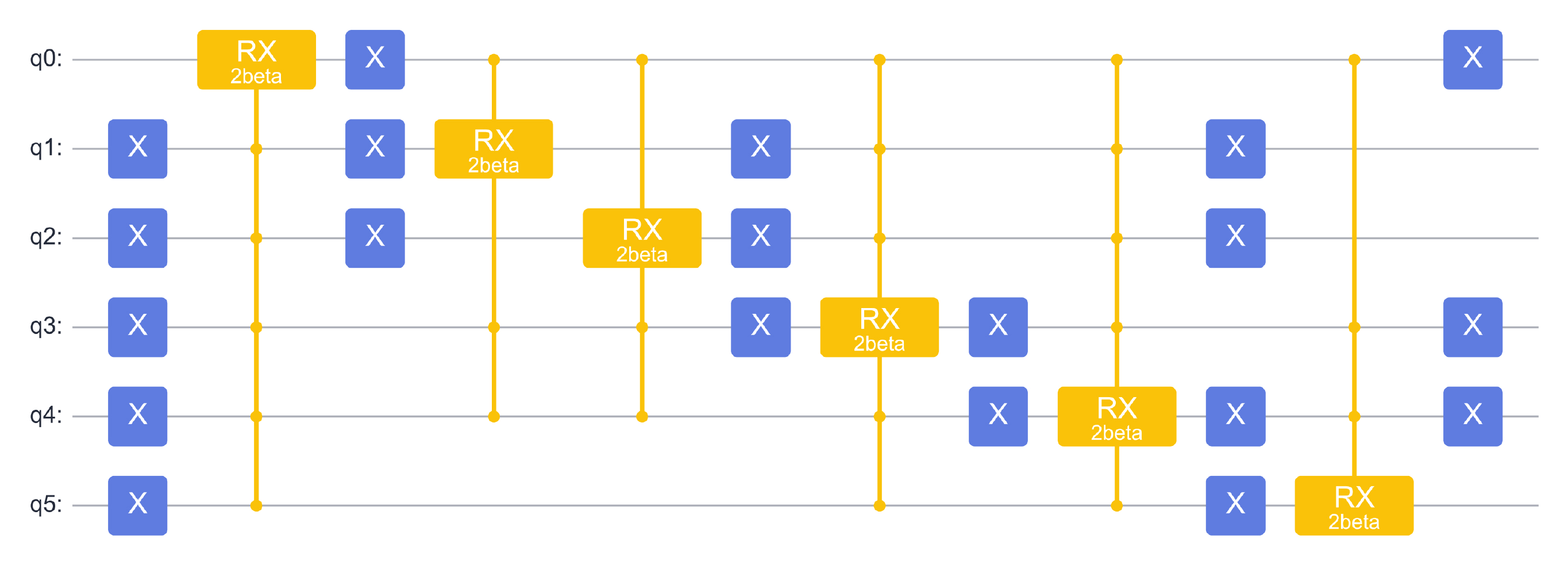}
    \caption{The new 1-level quantum circuit after removing the single-qubit rotating gates $R_Z$.}
\label{Fig.6}
\end{figure*}

\subsection{Quantum circuit optimization}
\label{Subsec:Evaluate}
 The QAOA+ mapping comprises phase-separation operators $U(H_P, \gamma)$,  mixing operators $U(H_M, \beta)$, and initial state $|x\rangle$ for MEC problem.  Based on  $H_M$ and $H_P$, the corresponding circuits of $U(H_M, \beta)$ are multiqubit-controlled-$R_X{(2\beta)}$ gates, and the corresponding circuits of $U(H_P, \gamma)$ are $n$ single-qubit rotating gates $R_Z$. For example: $X = \{1, \cdots , 12\}$ and $S = \{S_1, \cdots , S_6\}$, where $S_1 = \{1, 2, 4, 5, 6, 8, 9, 10\}$, $S_2 = \{1, 4, 6, 7\}$, $S_3 = \{5, 8, 9, 11, 12\}$, $S_4 = \{4, 7, 8, 9, 10, 11, 12\}$, $S_5 = \{2, 3, 4, 5, 6, 7, 11, 12\}$, $S_6 = \{3, 10, 12\}$. The corresponding figure is constructed, as shown in Fig. 4.

\renewcommand{\thefigure}{4}
\begin{figure}[htp]
    \centering
    \includegraphics[width=8cm]{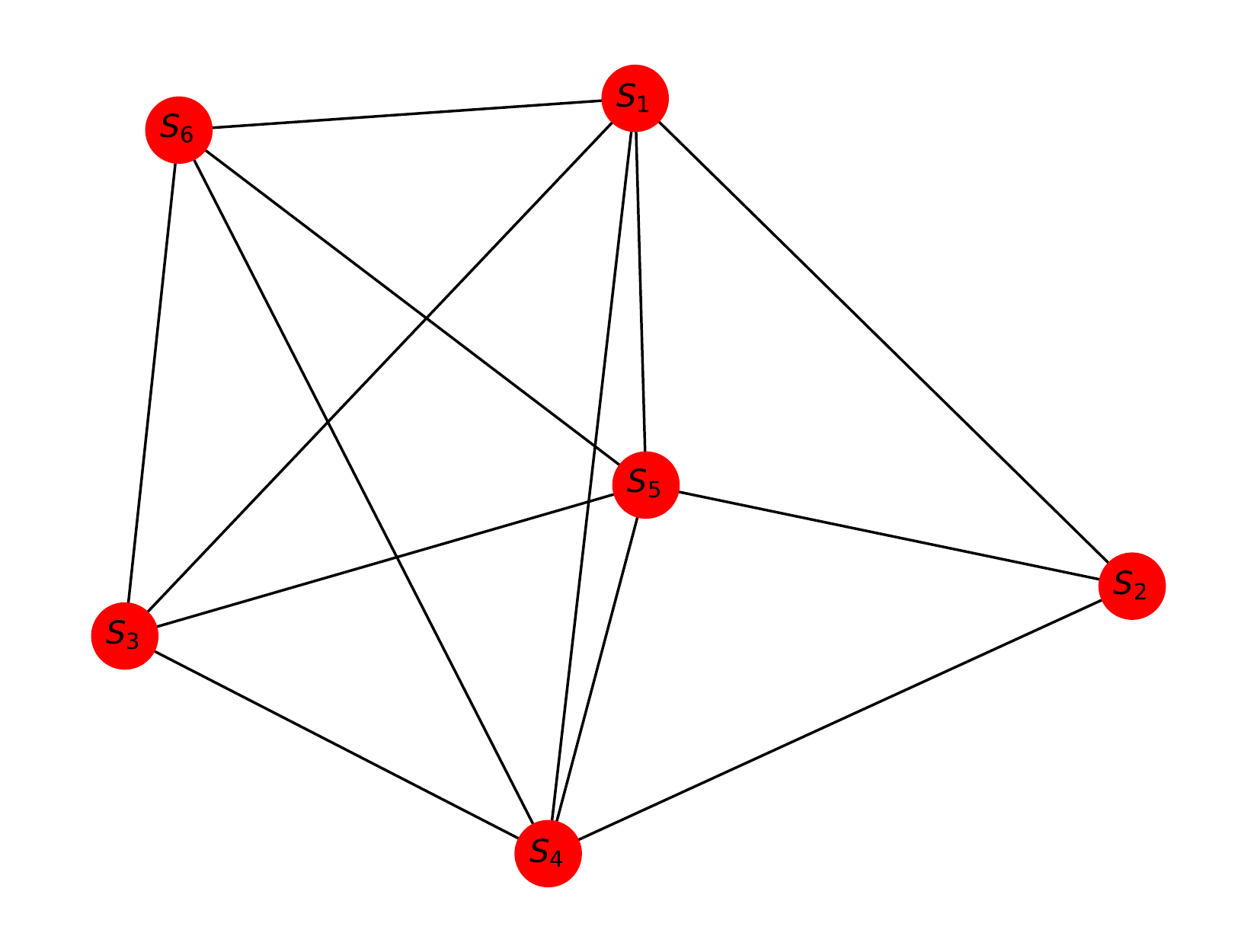}
    \caption{Graph representation of the instance.}
\label{Fig.4}
\end{figure}
\renewcommand{\thefigure}{7}
\begin{figure}[htp]
    \centering
    \includegraphics[width=8cm]{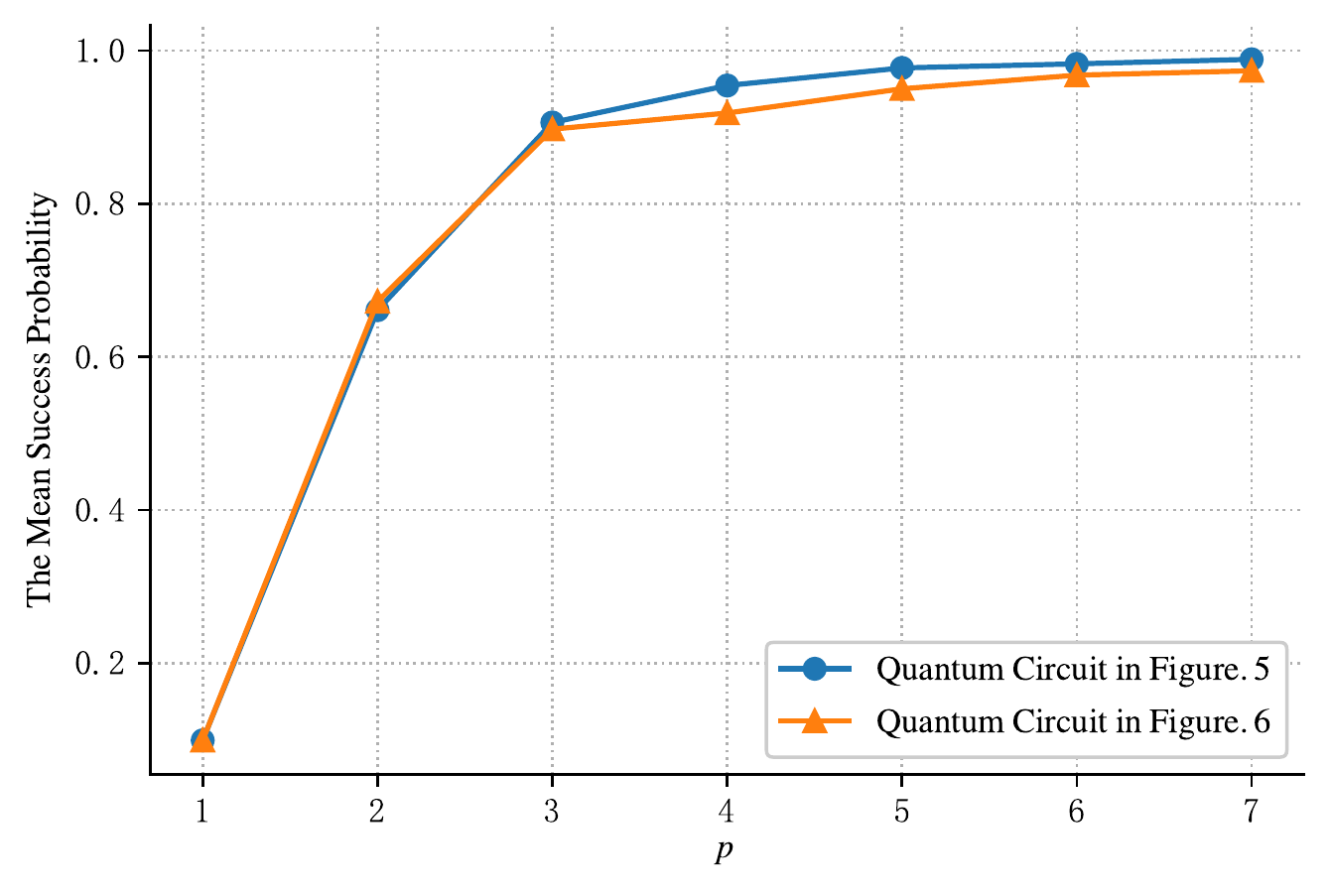}
    \caption{The comparison of mean success probability corresponding to these two circuits.}
\label{Fig.7}
\end{figure}

Based on discussion about weight $\lambda_i (i=1, 2)$, let $\lambda_2=\frac{1}{nm - 2}=1/70$, ${\lambda_1} = n \lambda_2 = 6/70$, the overall 1-level quantum circuit diagram of QAOA+ mapping is given, as shown in Fig. 5. Since single-qubit rotating gate $R_Z$ only changes the phase, we consider removing $n$ single-qubit rotating gates $R_Z$, as shown in Fig. 6. We find that this new $p$-level circuit only needs $p$ parameters, and the number of quantum gates is at least reduced by $np$. To study the ability of these two $p$-level circuits, the comparison of mean success probability is plotted using random initialization method in Figure. 7. The results show that $p$-level circuit only requires $p$ parameters, which can achieve an experimental effect similar to the original circuit with $2p$ parameters. And, the number of quantum gates is at least reduced by $np$ for $p$-level optimized circuit.\\

\section{SUMMARY AND PROSPECTS}
\label{Subsec:Dis}
\label{Sec:conclusion}
 To summarize, taking MEC problem as an example, we studied how to apply QAOA+ to NTFSP.  To find a trivial feasible solution, we transformed MEC into a multi-objective constrained optimization problem, where feasible space is composed of independent sets and the trivial solutions are easy to find. To solve the above multi-objective constrained optimization problem, we adopted the linear weighted sum method to construct target Hamiltonian. Finally, inspired by the parameter fixed strategy, we used this strategy to simulate instances with 6, 8, and 10 qubits. The numerical results show that the solution can be obtained with a probability close to 100\% for 6 and 8 qubits, even though level $p$ of the algorithm is low (see Fig.~\ref{Fig.3} for details). Besides, since single-qubit rotating gate $R_Z$ only changes the phase,  we optimized quantum circuit by removing the rotating gates $R_Z$. For $p$-level optimized circuit, the number of quantum gates is at least reduced by $np$ (see Fig.~\ref{Fig.6} for details). And, $p$-level optimized circuit only needs $p$ parameters, which can achieve an experimental effect similar to original circuit with $2p$ parameters (see Fig.~\ref{Fig.7} for details).

 In this work, we applied QAOA+ to solve the NTFSP, which provides a meaningful reference for how to use QAOA+ to solve such problems, and might greatly expand the application of the algorithm. In addition, our algorithm can also solve the tail-assignment problem with the minimum number of selected routes. See the detailed analysis in Appendix~\ref{Sec:The tail-assignment problem with the minimum number of selected routes}.

\begin{acknowledgments}
We thank Shijie Pan, Linchun Wan, and Xiumei Zhao for useful discussions on the subject. This work is supported by the Beijing Natural Science Foundation (Grant No. 4222031), the National Natural Science Foundation of China (Grant Nos. 61976024, 61972048, 62272056), and the 111 Project B21049.
\end{acknowledgments}

\begin{widetext}
\appendix
\section{The tail-assignment problem with the minimum number of selected routes}
\label{Sec:The tail-assignment problem with the minimum number of selected routes}
Airlines realize the optimal allocation of various production factors, through careful organization and production planning. Aircraft route allocation is an important part of the airline's organization and production planning. A reasonable and effective aircraft route allocation scheme helps to ensure the core utilization of the airline's resources, implement the airline's development strategy, ensure the safety of the airline's production and operation activities, and the realizability of the airline's revenue and expenditure budget in the current year. For a long time, operational research theory has been the source of innovation and development of international air transport industry, and has been widely used in the production and planning of various organizations of airlines, aviation revenue management and other fields. However, the airline network is one of the most complex networks in the world, and it is still developing rapidly. In the face of more and more complicated airline networks, the production and plans of airlines also pose great challenges to researchers.

The tail-assignment problem \cite{45,46} is an essential issue in the production planning of airlines. It is also one of the main contents in the operation control of airlines, where the goal is to decide which individual aircraft should operate which flight. By introducing the concept of route, the problem of aircraft-to-flight assignment is transformed into the problem of aircraft-to-route assignment. Each route starts and ends at the hub airport, and there are fixed departure and arrival times at the hub airport, thus reducing the scale and complexity of the problem. The tail-assignment problem is a combinatorial optimization problem essentially, which is NP-complete \cite{49}, and also a hot topic studied by scholars.

 Now, let $F$ denote the set of flights, and $R$ the set of all legal routes. Denote by $c_{r_i}$ the cost of route $r_i \in R$. Let $a_{fr_i}$ be 1 if flight $f$ is covered by route $r_i$ and 0 otherwise. The decision variable $x_{r_i}$ is 1 if route $r_i$ should be used in the solution, and 0 otherwise. The tail-assignment problem \cite{23,24,48} can now be formulated as
 \begin{align}
& min     \quad \sum_{i=1}^{ \left| R \right|} c_{r_i}x_{r_i}, \label{eqA1}\\
& s. t.  \quad  \sum_{r_i \in R} a_{fr_i}x_{r_i} = 1, \quad \forall f \in F,\label{eqA2}\\
& \quad \quad   x_{r_i} \in \{0, 1 \}, \quad \forall r_i \in R.
 \end{align}
The objective  Eq. (\ref{eqA1}) is to minimize the total cost of the selected routes, subject to constraints Eq. (\ref{eqA2}) ensuring that the set of routes in a solution should contain flight $f$ exactly once each flight. The model is an example of an exact cover problem, which is NP-complete \cite{49}.

According to the mathematical model of the MEC problem, the tail-assignment problem can also be expressed as the following
\begin{align}
& min    \quad  \sum_{i=1}^{ \left| R \right|} c_{r_i}x_{r_i}, \label{eqA4}\\
& max     \quad \sum_{i=1}^{ \left| R \right|} \omega_{r_i}x_{r_i}, \label{eqA5}\\
& s. t.  \quad x_{r_i} + x_{r_j} \le 1, \quad r_i \cap r_j \neq \emptyset, \label{eqA6}\\
& \quad \quad x_{r_i}, x_{r_j} \in \{0, 1 \},
\end{align}
where $c_{r_i}$ represents the cost of $r_i$, and $\omega_{r_i} = \left| r_i \right|$. The objective Eq. (\ref{eqA4}) is to minimize the total cost of the selected routes, and the objective  Eq. (\ref{eqA5}) is to maximize the sum of the weights of each route, subject to constraints  Eq. (\ref{eqA6}) ensuring that two routes with $r_i \cap r_j \neq \emptyset$ cannot be selected simultaneously.

In particular, we introduce a new target: the minimum number of aircraft (i.e., the minimum number of selected routes). The tail-assignment problem with the minimum number of selected routes can now be formulated as
\begin{align}
& min    \quad  \sum_{i=1}^{ \left| R \right|} c_{r_i}x_{r_i}, \label{eqA8}\\
& min    \quad  \sum_{i=1}^{ \left| R \right|} x_{r_i}, \label{eqA9}\\
& max     \quad \sum_{i=1}^{ \left| R \right|} \omega_{r_i}x_{r_i}, \label{eqA10}\\
& s. t.  \quad x_{r_i} + x_{r_j} \le 1, r_i \cap r_j \neq \emptyset, \label{eqA11}\\
& \quad \quad x_{r_i}, x_{r_j} \in \{0, 1 \}.
\end{align}

 The objective  Eq. (\ref{eqA8}) is to minimize the total cost of the selected routes, and the objective Eq. (\ref{eqA9}) is to minimize the total number of the selected routes, and the objective Eq. (\ref{eqA10}) is to maximize the sum of the weights of each route, subject to constraints  Eq. (\ref{eqA11}) ensuring that two routes with $r_i \cap r_j \neq \emptyset$ cannot be selected simultaneously. According to the importance of the objective function, it is arranged as  Eq. (\ref{eqA10}),  Eq. (\ref{eqA9}), and  Eq. (\ref{eqA8}) in descending order.

The tail-assignment problem with the minimum number of selected routes is transformed into a single objective constrained optimization problem
\begin{align}
& max     \quad \lambda_1\sum_{i=1}^{ \left| R \right|} \omega_{r_i}x_{r_i} - \lambda_2\sum_{i=1}^{ \left| R \right|} x_{r_i}- \lambda_3 \sum_{i=1}^{ \left| R \right|} c_{r_i}x_{r_i}, \\
& s. t.  \quad x_{r_i} + x_{r_j} \le 1, r_i \cap r_j \neq \emptyset, \\
& \quad \quad x_{r_i}, x_{r_j} \in \{0, 1 \},
\end{align}
where $\lambda_1 > \lambda_2 > \lambda_3 >0$, and their values can be determined according to experience. The corresponding phase separation Hamiltonian is obtained by replacing $x_{r_i}$ with $\frac{1-\sigma_i^Z}{2}$
\begin{align}
H_P = \lambda_1\sum_{i=1}^{ \left| R \right|} \omega_{r_i}\frac{1-\sigma_i^Z}{2} - \lambda_2\sum_{i=1}^{ \left| R \right|} \frac{1-\sigma_i^Z}{2}- \lambda_3 \sum_{i=1}^{ \left| R \right|} c_{r_i}\frac{1-\sigma_i^Z}{2}.
\end{align}

 We study instances for three different problem sizes of tail-assignment problem with the minimum number of selected routes given in Table II, corresponding to 6, 8, and 10 routes, respectively.
\begin{table}[!ht]
\caption{Information about the tail-assignment problem with the minimum number of selected routes instances.}
\begin{tabular}{|p{50pt}<{\centering}|p{50pt}<{\centering}|p{90pt}<{\centering}|p{90pt}<{\centering}|}
\hline
$| F |$ &  $| R |$ &  Number of instances & Number of solutions \\
\hline
12  & 6 & 1 & 1  \\
\hline
16  & 8 & 1 & 1  \\
\hline
20  & 10 & 1 & 1  \\
\hline
\end{tabular}
\end{table}

We conducted numerical simulations for the examples in the table shown in Fig. 8. In Fig. 8, the mean success probability as a function of  level $p$ for the three different problem sizes is plotted with random initialization method. The numerical result shows that the mean success probability of 10 route instance is higher than that of 8 route instance for $p\le 6$. This fact can seem counterintuitive, as one could naively think that larger instances correspond to harder problems. In addition, we note that the mean success probability of the 10 route instance shows a downward trend for $7\le p\le 8$. For the counterintuitive phenomena shown in Fig. 8, we will study them in future work.
\begin{figure}[htp]
    \centering
    \includegraphics[width=8.5cm]{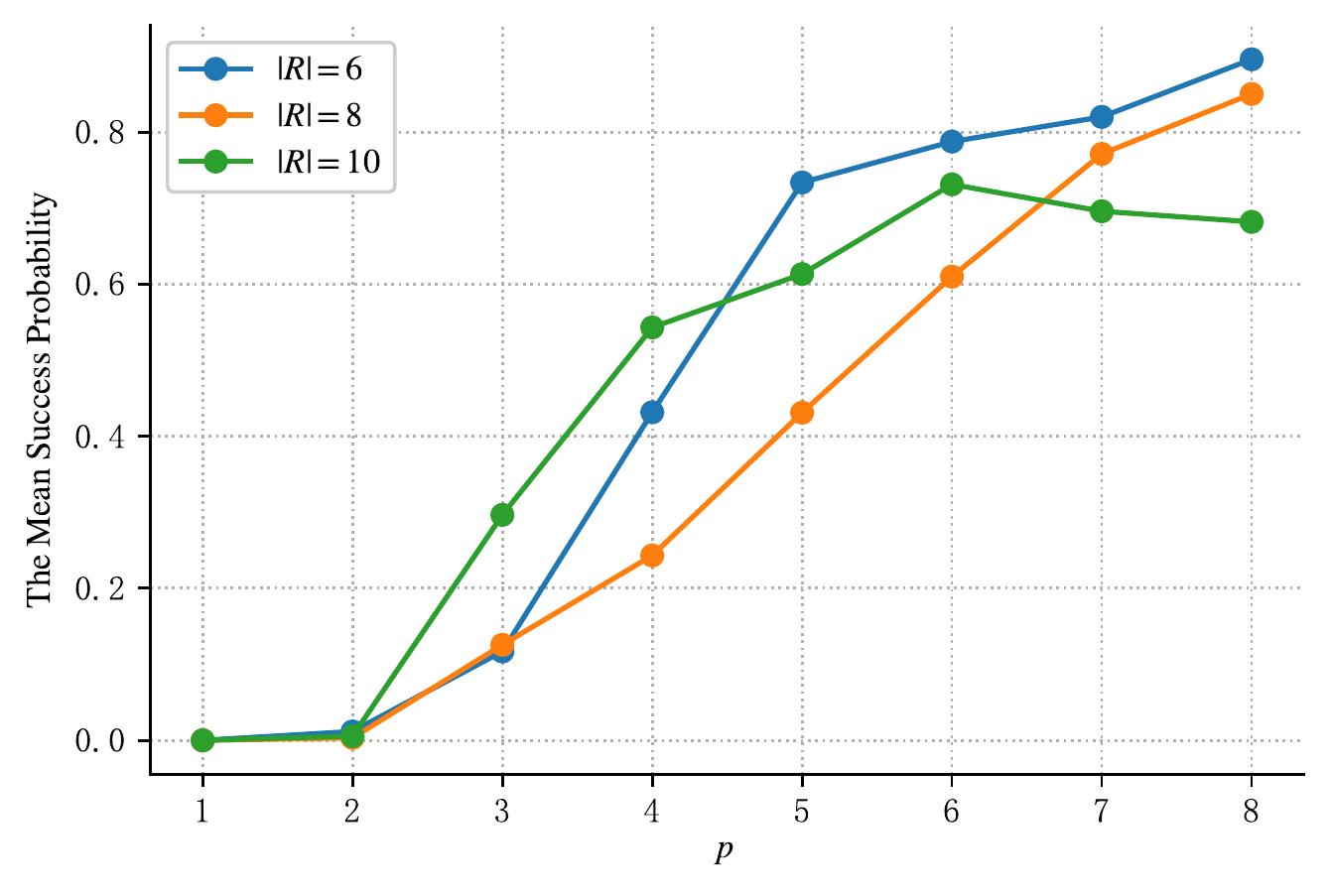}
    \caption{Success probability for solving the tail-assignment problem with the minimum number of selected routes.}
    \label{Fig.10}
\end{figure}
\end{widetext}
\bibliography{biblatex-phys}

\begin{thebibliography}{11}

\bibitem{1}
P. W. Shor, Algorithms for quantum computation: Discrete logarithms and factoring, in \emph{Proceedings 35th Annual Symposium on Foundations of Computer Science}, pp. 124-134 (1994).
\bibitem{2}
L. K. Grover, Quantum mechanics helps in searching for a needle in a haystack, Phys. Rev. Lett. \textbf{79} (2), 325-328 (1997).


\bibitem{4}
C. H. Yu, F. Gao, C. Liu, D. Huynh, M. Reynolds, and J. Wang, Quantum algorithm for visual tracking, Phys. Rev. A \textbf{99}, 022301 (2019).
\bibitem{5}
C. H. Yu, F. Gao, and Y. Q. Wen, An improved quantum algorithm for ridge regression, IEEE Trans. Know. Data Eng. \textbf{33} (3),  858-866 (2021).

\bibitem{7}
S. Lloyd, M. Mohseni, and P. Rebentrost, Quantum principal component analysis, Nature Physics \textbf{10}, 631 (2014).
\bibitem{8}
S. J. Pan, L. C. Wan, H. L. Liu, Q. L. Wang, S. J. Qin, Q. Y. Wen, and F. Gao, Improved quantum algorithm for A-optimal projection, Phys. Rev. A \textbf{102} (5), 052402 (2020).
\bibitem{9}
S. J. Pan, L. C. Wan, H. L. Liu, Y. S. Wu, S. J. Qin, Q. Y. Wen, and F. Gao, Quantum algorithm for neighborhood preserving embedding, Chinese Physics B \textbf{31}, 060304 (2022).
\bibitem{10}
Y. M. Li, H. L. Liu, S. J. Pan, S. J. Qin, F. Gao, and Q. Y. Wen, Quantum discriminative canonical correlation analysis, arXiv:2206.05526.

\bibitem{3}
H. Wang, Y. Xue, Y. Qu, et al., Multidimensional Bose quantum error correction based on neural network decoder, npj Quantum Inf \textbf{8}, 134 (2022).

\bibitem{11}
L. C. Wan, C. H. Yu, S. J. Pan, F. Gao, Q. Y. Wen, and S. J. Qin, Asymptotic quantum algorithm for the Toeplitz systems, Phys. Rev. A \textbf{97}, 062322 (2018).
\bibitem{12}
H. L. Liu, S. J. Qin, L. C. Wan, C. H. Yu, S. J. Pan, F. Gao, and Q. Y. Wen, A quantum algorithm for solving eigenproblem of the Laplacian matrix of a fully connected weighted graph, arXiv:2203.14451.
\bibitem{13}
L. C. Wan, C. H. Yu, S. J. Pan, S. J. Qin, F. Gao, and Q. Y. Wen, Block-encoding-based quantum algorithm for linear systems with displacement structures, Phys. Rev. A \textbf{104}, 062414 (2021).
\bibitem{14}
H. L. Liu, C. H. Yu, L. C. Wan, S. J. Qin, F. Gao, and Q. Y. Wen, Quantum mean centering for block-encoding-based quantum algorithm, Physica A: Statistical Mechanics and its Applications \textbf{607}, 128227 (2022).
\bibitem{15}
M. Guo, H. Liu, Y. Li, W. Li, F. Gao, S. Qin, and Q. Wen, Quantum algorithms for anomaly detection using amplitude estimation, Physica A: Statistical Mechanics and its Applications \textbf{604} (3), 127936 (2022).
\bibitem{16}
Z. Q. Li, B. B. Cai, H. W. Sun, H. L. Liu, L. C. Wan, S. J. Qin, Q. Y. Wen, and F. Gao, Novel quantum circuit implementation of Advanced Encryption Standard with low costs, Sci. China-Phys. Mech. Astron. \textbf{65}, 290311 (2022).


\bibitem{17}
E. Farhi, J. Goldstone, and S. Gutmann, A quantum approximate optimization algorithm, arXiv:1411.4028.
\bibitem{18}
P. Botsinis, D. Alanis, S. Feng, Z. Babar, H. V. Nguyen, D. Chandra, S. X. Ng, R. Zhang, and L. Hanzo, Quantum-assisted indoor localization for uplink mm-wave and downlink visible light communication systems, IEEE Access \textbf{5}, 23327-23351 (2017).
\bibitem{19}
Z. Wang, S. Hadfield, Z. Jiang, and E. G. Rieffel, Quantum approximate optimization algorithm for maxcut: A fermionic view, Physical Review A \textbf{97}, 022304 (2018).
\bibitem{20}
R. Herrman, L. Treffert, J. Ostrowski, P. C. Lotshaw, T. S. Humble, and G. Siopsis, Impact of graph structures for QAOA on maxcut, Quantum Information Processing \textbf{20} (9), 289 (2021).
\bibitem{21}
Y. J. Zhang, X. D. Mu, X. W. Liu, X. Y. Wang, X. Zhang, K. Li, T. Y. Wu, D. Zhao, and C. Dong, Applying the quantum approximate optimization algorithm to the minimum vertex cover problem, Applied Soft Computing, \textbf{118}, 108554 (2022).
\bibitem{22}
J. R. Weggemans, A. Urech, A. Rausch, R. Spreeuw, R. Boucherie, F. Schreck, K. Schoutens, J. Min\'{a}\v{r}, and F. Speelman, Solving correlation clustering with QAOA and a Rydberg qudit system: a full-stack approach, Quantum \textbf{6}, 687 (2022).


\bibitem{26}
R. Biswas, et al., A NASA perspective on quantum computing: Opportunities and challenges, Parallel Computing \textbf{64}, 81-98 (2017).
\bibitem{27}
E. G. Rieffel, D. Venturelli, B. O'Gorman, M. B. Do, E. M. Prystay, and V. N. Smelyanskiy, A case study in programming a quantum annealer for hard operational planning problems, Quantum Information Processing, \textbf{14} (1), 1-36 (2015).
\bibitem{28}
S. Hadfield, On the representation of Boolean and real functions as Hamiltonians for quantum computing, arXiv:1804.09130.
\bibitem{29}
I. Hen and F. M. Spedalieri, Quantum Annealing for Constrained Optimization, Phys. Rev. Applied \textbf{5} (3), 034007 (2016).
\bibitem{30}
V. Choi, Different adiabatic quantum optimization algorithms for the NP-complete exact cover and 3SAT problems, arXiv:1010.1221.
\bibitem{31}
S. Hadfield, Z. Wang, B. O'Gorman, E. G. Rieffel, D. Venturelli, and R. Biswas, From the quantum approximate optimization algorithm to a quantum alternating operator ansatz, Algorithms \textbf{12} (2), 34 (2019).

\bibitem{33}
Z. H. Wang, N. C. Rubin, J. M. Dominy, and E. G. Rieffel, XY mixers: Analytical and numerical results for the quantum alternating operator ansatz, Phys. Rev. A \textbf{10} (1), 012320 (2020).

\bibitem{34}
Z. H. Saleem, Max-independent set and the quantum alternating operator ansatz, International Journal of Quantum Information \textbf{18} (4), 2050011 (2020).
\bibitem{35}
J. Cook, S. Eidenbenz, and A. Brtschi, The Quantum Alternating Operator Ansatz on Max-$k$ Vertex Cover, in \emph{APS March Meeting}, \textbf{65}, 1 (2020).
\bibitem{36}
S. Chatterjee and D. Bera, Applying the Quantum Alternating Operator Ansatz to the Graph Matching Problem, arXiv.2011.11918.
\bibitem{37}
M. Fingerhuth, B. Tom\'{a}, and C. Ing, A quantum alternating operator ansatz with hard and soft constraints for lattice protein folding, arXiv.1810.13411.

\bibitem{23}
P. Vikst$\mathring{a}$l, M. Grnkvist, M. Svensson, M. Andersson, G. Johansson, and G. Ferrini, Applying the Quantum Approximate Optimization Algorithm to the Tail-Assignment Problem, Phys. Rev. Applied \textbf{14} (3), 034009 (2020).
\bibitem{24}
M. Svensson, M. Andersson, M. Grnkvist, P. Vikst$\mathring{a}$l, D. Dubhashi, G. Ferrini, and G. Johansson, A Heuristic Method to solve large-scale Integer Linear Programs by combining Branch-and-Price with a Quantum Algorithm, arXiv:2103.15433.
\bibitem{25}
A. Lucas, Ising formulations of many NP problems, Front. Phys. \textbf{2}, 5 (2014).

\bibitem{38}
I. P. Stanimirovi\'{c}, M. L. Zlatanovi\'{c}, and M. D. Petkovi\'{c}, On the linear weighted sum method for multi-objective optimization, Facta Univ. Ser. Math. Inform. \textbf{26}, 49-63 (2011).
\bibitem{39}
L. Zhou, S. T. Wang, S. Choi, H. Pichler, and M. D. Lukin, Quantum approximate optimization algorithm: Performance, mechanism, and implementation on near-term devices, Phys. Rev. X \textbf{10} (2), 021067 (2020).
\bibitem{40}
C. G. Broyden, The Convergence of a Class of Double-rank Minimization Algorithms 1. General Considerations, IMA Journal of Applied Mathematics \textbf{6}, 76 (1970).
\bibitem{41}
R. Fletcher, A new approach to variable metric algorithms, The Computer Journal \textbf{13}, 317 (1970).
\bibitem{42}
D. Goldfarb, A family of variable-metric methods derived by variational means, Mathematics of Computation \textbf{24}, 23 (1970).
\bibitem{43}
D. F. Shanno, Conditioning of quasi-Newton methods for function minimization, Mathematics of Computation \textbf{24}, 647 (1970).
\bibitem{44}
X. Lee, Y. Saito, D. Cai, and N. Asai, Parameters Fixing Strategy for Quantum Approximate Optimization Algorithm, in \emph{2021 IEEE International Conference on Quantum Computing and Engineering}, pp. 10-16 (2021).

\bibitem{45}
 M. Gr\"{o}nkvist, The Tail Assignment Problem, Ph.D. thesis, Chalmers University of Technology and Gteborg University (2005).
\bibitem{46}
T. L. Jacobs, L. A. Garrow, M. Lohatepanont, Frank S. Koppelman, G. M. Coldren, and H. W. Purnomo, Airline Planning and Schedule Development, in \emph{Quantitative Problem Solving Methods in the Airline Industry}, pp. 35-39 (2012).
\bibitem{47}
M. Gr\"{o}kvist and J. Kjerrstrm, Tail Assignment in Practice, in \emph{Operations Research Proceedings 2004}, pp. 166-173 (2004).

\bibitem{48}
L. Martins, Applying Quantum Annealing to the Tail Assignment Problem, Ph.D. thesis, University of Porto (2020).
\bibitem{49}
R. M. Karp, Reducibility Among Combinatorial Problems, in \emph{Complexity of Computer Computations}, pp. 85-103 (1972).

\bibitem{51}
MindQuantum Developer, MindQuantum, version 0.6.0, https://gitee.com/mindspore/mindquantum (2021).

\end{thebibliography}
\end{document}